\documentclass[12pt]{article}
\pdfoutput=1
\usepackage{epsfig}
\usepackage{amsmath}
\usepackage{amssymb}
\usepackage{color}
\usepackage{sectsty}

\usepackage{bm}
\usepackage{sidecap}
\usepackage[round]{natbib}
\usepackage{doi}

\usepackage{orcidlink}  % for orcid ID
\hypersetup{%  This removes the blue box around the ID label
    pdfborder = {0 0 0}
}

\def\rovariation{
\begin{figure*}[b!]
\includegraphics[trim=0 0 50 0, width=0.9\textwidth, keepaspectratio=true]{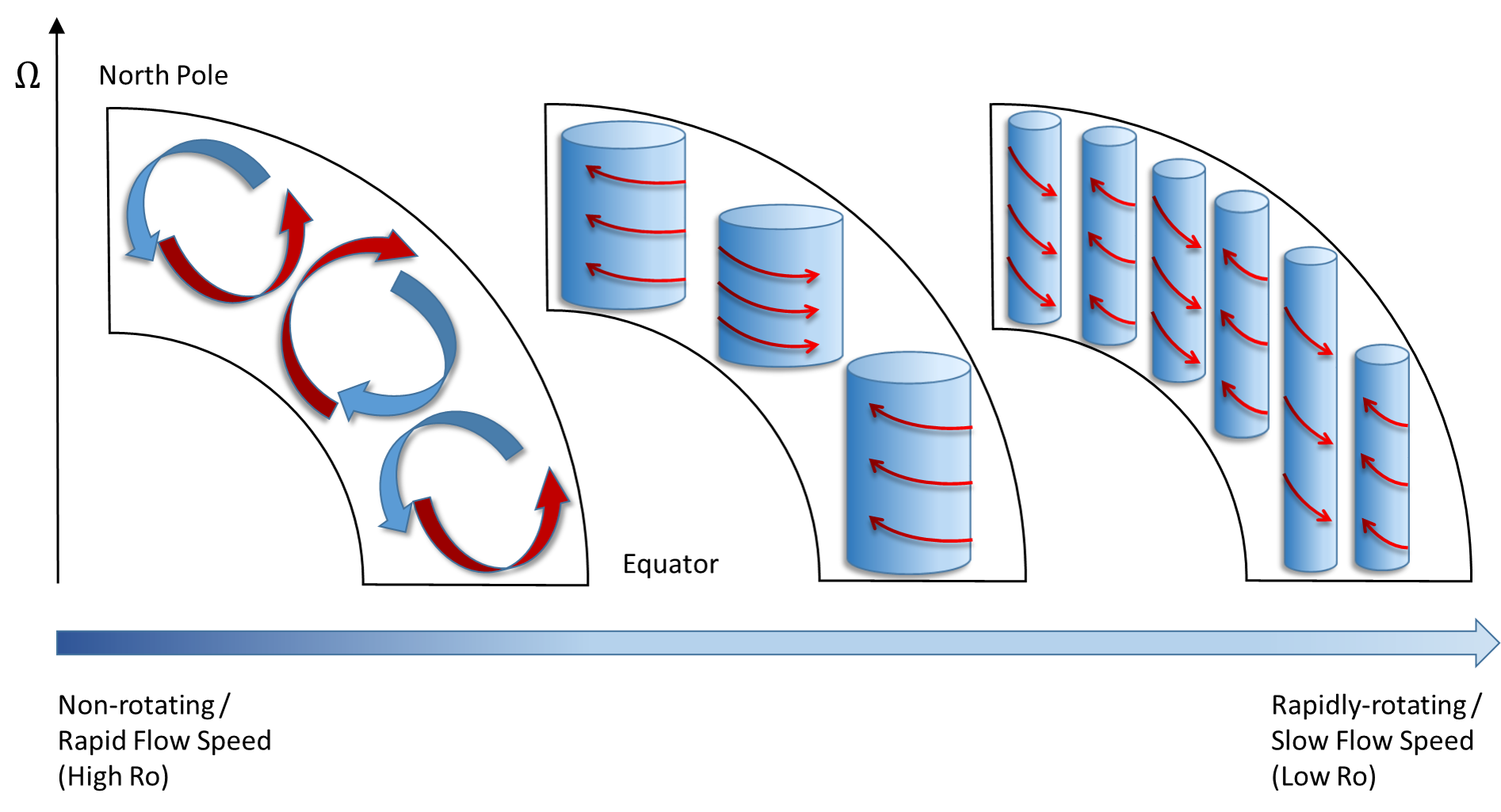}
\centerline{\epsfig{file=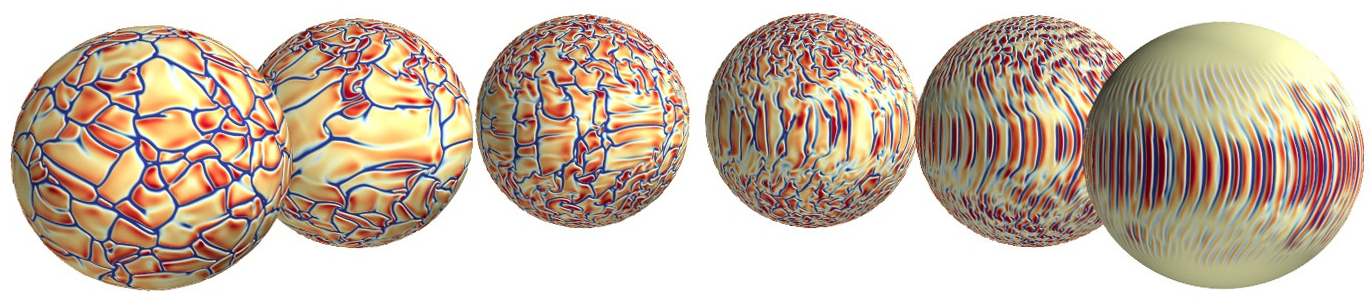,width=0.9\textwidth}}
\vspace{-.2in}
\caption{\footnotesize\label{fig:rovariation}  Rotational influence on convective structure.  (\textit{upper}) Schematic convective flow structures as realized in rotating convection, viewed in cutaway of northern hemisphere.  (\textit{lower}) 3-D rendering of simulated solar-like convection realized under different Rossby numbers.   Regions of upflow (downflow) are rendered in red (blue).  As rotational influence increases from left (high Ro) to right (low Ro), convective patterns become increasingly more helical and columnar \citep{Featherstone2016}.}
\end{figure*}
}

\def\dynamostates{
\begin{figure*}[t]
\centering
\vspace{-0.45in}
\includegraphics[width=\textwidth, keepaspectratio=true]{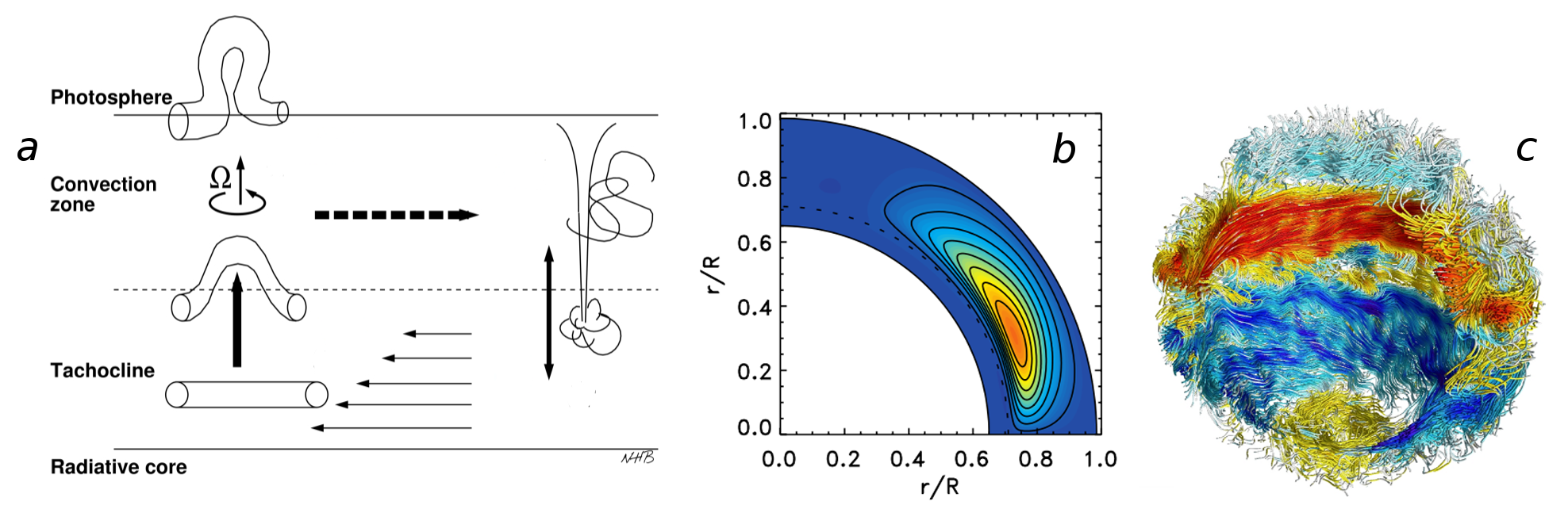}
\vspace{-0.35in}
\caption{\footnotesize Potential solar-dynamo mechanisms. (\textit{a}) An interface dynamo, sustained by turbulent pumping of magnetic field into the tachocline.   (\textit{b}) Meridional flow profile employed in a flux-transport dynamo whose cycle timing is tied to the circulatory timescale (black streamlines overlying colored streamfunction; \citet{Rempel2006}).  (\textit{c}) Magnetic field lines in  a ``wreathy'' dynamo, realized in low-Ro convection without a tachocline. Red/blue coloring indicates opposite signs of toroidal-field polarity \citep{Brown2011}.}
\label{fig:dynamostates}
\end{figure*}
}

\def\mcvariation{
\begin{figure*}[t]
\centering
\vspace{-0.45in}
\includegraphics[trim=20 100 0 0, width=\textwidth, keepaspectratio=true]{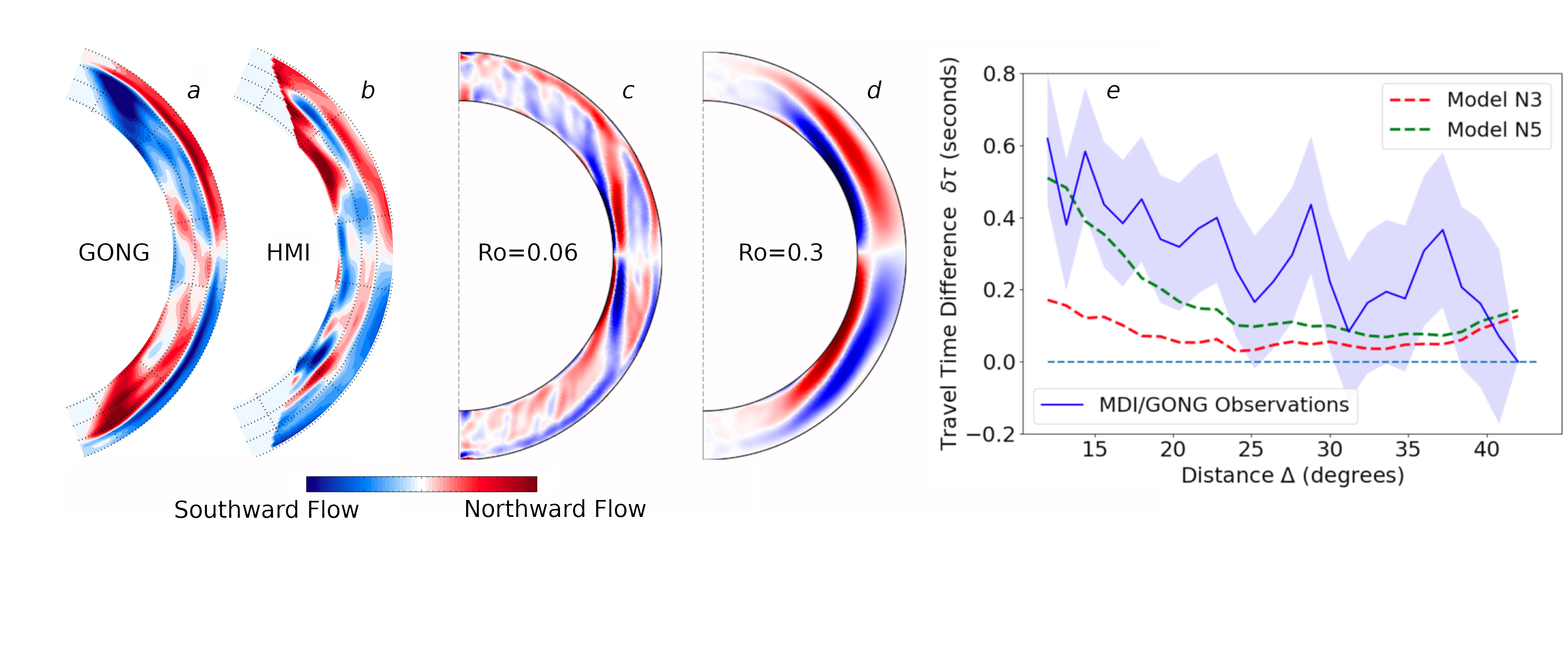}
\caption{\footnotesize  Continuing questions regarding deep meridional flow. (\textit{a},\textit{b}) Meridional flow as obtained from GONG and HMI observations, showing the latitudinal component of the flow, averaged over time and longitude; adapted from \citet{Jackiewicz2015}. (\textit{c},\textit{d})  Similar rendering of possible flow profiles as realized in solar convection simulations run with different Rossby numbers (indicated; adapted from the survey of \citet{Hindman2020}). (\textit{e}) Comparison between the travel times obtained from observations and from two simulated meridional flow profiles  \citep[denoted by N3 and N5;][]{stejko2022}.   Discrepancies between observations and the lack of polar flow data make it difficult to distinguish between possible flow regimes.  Models and observations are nevertheless now making meaningful contact, as illustrated in $(e)$. }
\label{fig:mcvariation}
\vspace{-0.15in}
\end{figure*}
}

\def\polar{
\begin{figure*}[t]
\centering
\vspace{-0.35in}
\includegraphics[trim=0 15 0 0, width=\textwidth, keepaspectratio=true]{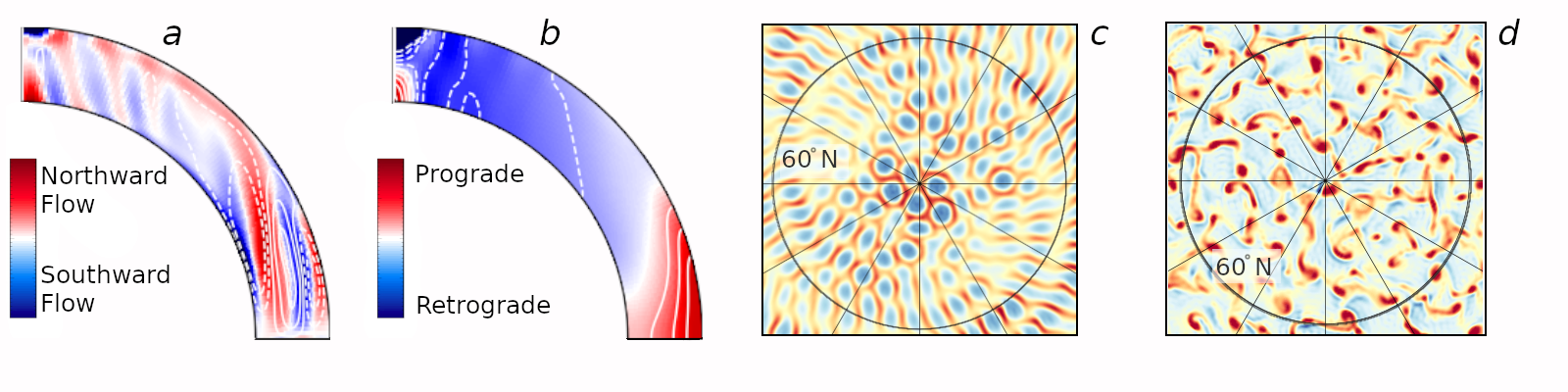}
\caption{\footnotesize Polar clues to interior convective structure as suggested by simulation \citep[adapted from][]{Hindman2020}.  (\textit{a}) Meridional flow realized at Ro=0.05.  At such low Ro, countercells develop in the polar regions (\textit{b}) Accompanying differential rotation. In the polar region, there is increased spin-down of near-surface layers and spin-up of deep CZ.  At larger values of Ro, prograde differential rotation can develop at all depths in the polar region. (\textit{c}) Polar flow structures for Ro=0.01 as viewed from the pole.  Reference circle indicates latitude 60$^\circ$.  Regions of downflow are indicated in red and upflow in blue.  (\textit{d}) Similar view of polar convection, but for Ro=0.03.   When viewed from the poles, both mean and convective flows can provide insight into deep interior convection.}
\label{fig:polar_flow}
\vspace{-0.15in}
\end{figure*}
}

\newcommand\Beq{\begin{eqnarray}} 
\newcommand\Eeq{\end{eqnarray}}

\newcommand{\SWRI}{\noindent $^1$ Southwest Research Institute}
\newcommand{\NWU}{\noindent $^2$ Northwestern University, Evanston, IL}
\newcommand{\UCLA}{\noindent $^3$ University of California, Los Angeles}
\newcommand{\CUB}{\noindent $^4$ University of Colorado, Boulder}
\newcommand{\UCSC}{\noindent $^5$ University of California, Santa Cruz}
\newcommand{\MIT}{\noindent $^6$ Massachusetts Institute of Technology}
\newcommand{\NCAR}{\noindent $^7$ National Center for Atmospheric Research}
\newcommand{\UFMG}{\noindent $^8$ Universidade Federal de Minas Gerais}
\newcommand{\CHICO}{\noindent $^9$ California State University, Chico}
\newcommand{\BATES}{\noindent $^{10}$ Bates College, Lewiston, ME}
\newcommand{\AUSTIN}{\noindent $^{11}$ University of Texas at Austin}
\newcommand{\NJIT}{\noindent $^{12}$ New Jersey Institute of Technology}
\newcommand{\UE}{\noindent $^{13}$ University of Edinburgh, School of Maths}

\newcommand{\AMES}{\noindent $^{14}$ NASA Ames Research Center}
\newcommand{\NOAA}{\noindent $^{15}$ NOAA Space Weather Prediction Center}

\newcommand{\coauthor}[3]{#1$^{#3}$\orcidlink{#2}}

\sectionfont{\fontsize{13}{13}\selectfont}
\subsectionfont{\fontsize{13}{13}\selectfont}

\begin{document}

\thispagestyle{empty}
\noindent
\begin{center}
{\bf\large The Puzzling Structure of Solar Convection:\\  Window into the Dynamo}
\end{center}
\medskip
%\orcidlink{0000-0002-1825-0097}
\noindent\textbf{Primary Author: }\coauthor{Nicholas A. Featherstone}{0000-0002-2256-5884}{1}\smallskip
\\
\noindent\textbf{Co-Authors: } \coauthor{Evan H. Anders}{0000-0002-3433-4733}{2},
\coauthor{Kyle C. Augustson}{0000-0003-4714-1743}{2},
\coauthor{Jonathan M. Aurnou}{0000-0002-8642-2962}{3},\\
\coauthor{Catherine Blume}{0000-0001-9004-5963}{4}, 	
\coauthor{Benjamin P. Brown}{0000-0001-8935-219X}{4},
\coauthor{Nicholas Brummell}{0000-0003-4350-5183}{5},
\coauthor{Keaton J. Burns}{0000-0003-4761-4766}{6},\\
\coauthor{Michael A. Calkins}{0000-0002-6917-5365}{4},
\coauthor{Maria Camisassa}{0000-0002-3524-190X}{4},
\coauthor{Mausumi Dikpati}{0000-0002-2227-0488}{7},
\coauthor{Yuhong Fan}{0000-0003-1027-0795}{7}
\coauthor{J.R. Fuentes}{0000-0003-2124-9764}{4},
\coauthor{Gustavo Guerrero}{0000-0002-2671-8796}{8},
\coauthor{Bradley W. Hindman}{0000-0001-7612-6628}{4},
\coauthor{Keith Julien}{0000-0002-4409-7022}{4},
\coauthor{Irina N. Kitiashvili}{0000-0003-4144-2270}{14},\\
\coauthor{Lydia Korre}{0000-0002-0963-4881}{4},
\coauthor{Daniel Lecoanet}{0000-0002-7635-9728}{2},
\coauthor{Bhishek Manek}{0000-0002-2244-5436}{4},
\coauthor{Loren Matilsky}{0000-0001-9001-6118}{5},
\coauthor{Mark Miesch}{0000-0003-1976-0811}{4,15},
\coauthor{Nicholas J. Nelson}{0000-0002-4967-5258}{9},
\coauthor{Jeffrey S. Oishi}{0000-0001-8531-6570}{10},
\coauthor{Whitney T. Powers}{0000-0001-7011-8029}{4},
\coauthor{Matthias Rempel}{0000-0001-5850-3119}{7}, \\
\coauthor{Krista Soderlund}{0000-0002-7901-3239}{11},
\coauthor{Andrey M. Stejko}{0000-0001-7483-3257}{12},
\coauthor{Geoffrey M. Vasil}{0000-0002-8902-5030}{13}
\medskip
\begin{center}
{\small
    \begin{tabular}{ l l }
    \SWRI & \NWU \\
    \UCLA & \CUB \\
    \UCSC & \MIT \\
    \NCAR & \UFMG \\
    \CHICO & \BATES \\
    \AUSTIN &  \NJIT \\ 
    \UE & \AMES \\
    \NOAA & 
    \end{tabular}
    }
\end{center}

\medskip
\noindent\textbf{Synopsis}\\
The operation of the solar dynamo, with all of its remarkable spatio-temporal ordering, remains an outstanding problem of modern solar physics.   A number of mechanisms that might plausibly contribute to its operation have been proposed, but the relative role played by each remains unclear.  This uncertainty stems from continuing questions concerning the speed and structure of deep-seated convective flows.  Those flows are in-turn thought to sustain both the Sun's turbulent EMF and the large-scale flows of differential rotation and meridional circulation suspected of influencing the dynamo's organization and timing.    

Ultimately, the convective and large-scale flow structure derive from the Coriolis force.  When the Coriolis effect is weak (rapid convective flow), convection exhibits little organization in its spatial structure, and the meridional flow is expected to assume a single-celled profile within each hemisphere.  Convection subject to strong Coriolis forces instead organizes into compact, spiraling columnar structures and sustains meridional flows that possess multiple cells in latitude and radius.  The implications for the dynamo in these two regimes are substantial, impacting the flux-transport properties of any assumed meridional flow and the convectively-driven EMF.

Continued progress in this area is complicated by (i) inconsistencies between helioseismic measurements of convective and meridional flow made with different techniques and instruments, and (ii) a lack of high-latitude data for convection, differential rotation, and meridional flow.   We suggest that the path forward to resolving these difficulties is twofold.  First, the acquisition of long-term helioseismic and emissivity measurements obtained from a polar vantage point is vital to complete our picture of the Sun's outer convection zone.  Second, sustained and expanded investment in theory-oriented and combined theory/observational research initiatives will be crucial to fully exploit these new observations and to resolve inconsistencies between existing measurements.

\clearpage

%\tableofcontents
%\contentsline{section}{test}{36}

\newpage
\pagenumbering{arabic}

%https://www.overleaf.com/project/62fec2afa510631c290f99cb
\rovariation

\section{Introduction}
An outstanding gap persists in our understanding of the solar dynamo.  Numerous and potentially viable descriptions of this physical system now exist, but not one has been definitively tied to the dynamics at work in the Sun, and for good reason.  Convection, an indispensable component of the dynamo, occurs in the presence of rotation; yet, owing to persistent questions surrounding the convective flow speed and structure at depth, the degree to which rotation influences convection remains largely unquantified throughout the solar interior.  

In a rotating, convecting system such as the Sun, it is the interplay between the Coriolis force and buoyant driving that ultimately shapes the convective structure and amplitude, and in turn,  the nature of the resulting dynamo.  The relative importance of these two effects is typically quantified through a ratio known as the Rossby number (Ro),
\begin{equation}
\label{eq:ro}
\mathrm{Ro} = \frac{\mathrm{rotation~period}}{\mathrm{convective~timescale}}.
\end{equation}
When the convective flow speed is high, and the resulting Ro is large, convective motions are largely unaffected by the Coriolis effect.  In this limit, fluid motions span the convection zone and exhibit little, if any apparent organization in their structure \citep[e.g.,][]{Ahlers2009}. When Ro is small, as happens when the convective turnover time exceeds a rotation period, the Coriolis effect is strong enough to induce significant curvature in the fluid trajectories \citep{Busse2002}.  Convective flows ultimately organize into spiraling, columnar structures with cross-sectional area proportional to Ro.  This situation is illustrated schematically and via simulation results in Fig. \ref{fig:rovariation}.

The Sun's location along the spectrum of behavior shown in Fig. 1 must profoundly impact key components of the solar dynamo.  Its differential rotation, meridional circulation, and turbulent EMF ultimately derive from the morphology and speed of its convecting plasma.  The latter is apparent upon consideration of the MHD induction term, namely
\begin{equation}
\label{eq:induction}
\frac{\partial \bm{B}}{\partial t} = \bm{\nabla}\times\left(\bm{v}\times \bm{B}\right),
\end{equation}
where the amplitude of the velocity vector $\bm{v}$ directly determines the rate of change in the magnetic field vector $\bm{B}$.  In turn, the structure of $\bm{v}$, via the spatial variation of its correlations with the magnetic field, is central to the inductive process.

\dynamostates

\subsection{The Multi-faceted Dynamo}

Several potential dynamo mechanisms may be operating in the solar interior, whether in isolation or in concert, and their relative importance hinges to a large degree on Ro.   Perhaps the most well-known of these is the interface dynamo mechanism \citep{Parker1993} which depends on the tachocline located at the base of the convection zone. The rotational shear characteristic of that region may serve to stretch poloidal field, pumped there by convective downdrafts, into organized toroidal field (Fig. \ref{fig:dynamostates}$a$).   As a result, the tachocline has been the presumed seat of the solar dynamo for some time, figuring prominently in the flux-transport class of dynamo models \citep[e.g., Fig. \ref{fig:dynamostates}$b$;][]{Dikpati1999,Rempel2006}.

The Sun's meridional circulation also plays a key role in flux-transport models, serving to advect toroidal flux near the base of the convection zone and resulting in the equatorward migration of active bands at the solar surface \citep[e.g.,][]{Charbonneau2020}.  By regulating this migration, the meridional flow also regulates the magnetic cycle period. %(though other factors, such as turbulent transport, can also contribute).

%\mcvariation

%The profile chosen is often motivated by simplicity, with a single cell of circulation in each hemisphere, and an equatorward flow of 2--3 m s$^{-1}$ near the base of the convection zone, yielding a cycle length of 11 years

Flux-transport models of the dynamo face complications, however, in that the structure and amplitude of deep meridional circulation remains a free parameter.  In the absence of observational constraints, a single-celled circulation profile within each hemisphere, consistent with near-surface observations of the meridional flow, has often been chosen  \citep[e.g.,][]{Dikpati1999}.   Whether or not this choice is reasonable depends on the Rossby number characteristic of deep convection, which is still a matter of debate.  

Computational studies of rotating convection have shown that the Rossby number is important in determining meridional flow structure \citep{Gastine2013,Guerrero2013,Gastine2014,Featherstone2015,Camisassa2022}.  Slow, low-Ro convection tends to generate a meridional circulation with multiple cells in radius (Fig.~\ref{fig:mcvariation}$c$).  At sufficiently low Ro, additional cells appear in latitude near the poles.  In the high-Ro regime, rapidly overturning convection generates a mono-cellular meridional circulation within each hemisphere (Fig.~\ref{fig:mcvariation}$d$). 

An interesting alternative to the interface-dynamo paradigm has arisen in recent years as fully nonlinear, 3-D dynamo models have begun to produce so-called ``wreathy'' magnetic fields.  These large-scale bands of toroidal magnetic field that encircle the convection zone are particularly interesting because can be generated either in the presence \textit{or in the absence} of a tachocline \citep[e.g., Fig. \ref{fig:dynamostates}$c$; ][]{Brown2010}. Such dynamos can undergo cycles where the toroidal belts modulate their amplitude, migrate in latitude, and change sign -- all key, observed characteristics of the solar dynamo \citep{Ghizaru2010,Brown2011,Racine2011,Augustson2015,Matilsky2020b}.  

Simulations resulting in wreathy behavior all have one property in common: the  Rossby number is small (typically between 0.01 and 0.03).  If the Sun is operating in such a low-Ro regime, the relative importance of the tachocline and meridional flow to the solar cycle becomes less clear.  \textbf{Without strict observational bounds on the Sun's Ro, we cannot assess either the likelihood of a wreathy dynamo, the importance of deep flux transport, or the role played by the tachocline.}  As a result, the predictive capability of any dynamo model is severely diminished.

%\mcvariation

\subsection{The Challenge}

The fundamental challenge inhibiting progress on this problem is that current measurements provide only limited insight into where the Sun lies on the spectrum of behavior illustrated in Fig. \ref{fig:rovariation}.    Solar convection-zone models that produce differential rotation resembling that of the Sun tend to possess substantial convective power on spatial scales similar to the convection zone depth, roughly consistent with the central images in Fig. \ref{fig:rovariation}.  This is not born out by observations, however.  Photospheric convective power instead peaks at the spatial scale of supergranulation, which is one-tenth the expected scale \citep[e.g.,][]{Hart1956,Leighton1962,Rincon2018}.  Larger-scale photospheric flows are much weaker and tend to be dominated by inertial waves \citep{Hathaway2021,Gizon2021}.  

\mcvariation

Below the photosphere, local helioseismology provides the potential to gain additional clues.  Local helioseismic analysis becomes challenging beyond a depth of about 30 Mm, however.   Helioseismic attempts to measure convective flows below this region remain inconclusive; some indicate substantial large-scale convective power, and others a lack thereof  \citep{Hanasoge2012,Greer2015,Proxauf2021}.   
%The time-distance measurements of \citep{Hanasoge2012} suggest that convection on spatial scales larger than 70 Mm can be at most 5--6 m s$^{-1}$ in the deep convection zone, implying that Ro$\le 0.01$.  When the same methods are applied to shallower layers, however, the inferred convective amplitudes disagree with those resulting from alternative, ring-diagram measurements by roughly an order of magnitude \citep{Greer2015,Proxauf2021}.

%Below the near-surface region, helioseismic procedures are most sensitive to the axisymmetric flows of differential rotation and, recently, meridional circulation.  
In principle, the Sun's meridional circulation can provide important clues into the nature of the underlying convection via the structural changes illustrated in Figs. \ref{fig:mcvariation}\textit{c,d}.  Unfortunately, there is presently no observational consensus concerning the structure of deep meridional circulation.  Local helioseismic analyses based on HMI data seem to suggest that the location of the circulation's return flow is relatively shallow and hint at multiple cells in radius \citep{Schad2012,Zhao2013,Chen2017}.  However, measurements deriving from ground-based GONG and space-based MDI data have failed to confirm this multi-cellular structure \citep[Figs. \ref{fig:mcvariation}\textit{a,b};][]{Jackiewicz2015,Gizon2020}.

%\polar

\section{Looking Forward}

%Continued progress on the dynamo problem will require the acquisition of new observations and the development of new techniques for imaging deep solar flows from existing data.  Moreover, the richness of dynamical behavior exhibited by even the limited parameter-space surveys embodied by Figs.\ref{fig:rovariation} and \ref{fig:mcvariation} suggests that a robust theoretical framework will be required to support the interpretation of those measurements.

Ultimately, it is novel helioseismic analysis, supported by a robust theoretical/modeling framework that promises to bridge outstanding gaps in our understanding of solar convection and the dynamo it sustains.  The promise of such a combined approach is well-illustrated by the recent
meridional circulation study carried out in \citet{stejko2022}.  By modeling the propagation of acoustic waves through simulated meridional flow profiles, those authors were able to generate synthetic time-distance measurements and compare them against actual measurements of the Sun.  Such comparisons can in turn be used to discriminate between potential models of interior convection, as illustrated in Fig. \ref{fig:mcvariation}$e$.   

The recent identification of inertial modes in the solar photosphere provides another example.  As these wave modes stem from the Coriolis force, their properties are sensitive to the underlying convective flow.   When combined with numerical simulations of deep solar convection, this rapidly-developing branch of helioseismology promises to help further discriminate between models \citep{Gizon2021,Bekki2022}.  

With these recent advances in mind, we recommend substantial investment in the areas of interior theory, modeling, and helioseismic analysis for the coming decade.  We believe this requires the investment both in broad projects that facilitate the union of the these thrusts, as well as investment in smaller, focused projects that enhance capabilities within these vital areas of research.  To that end, we advocate for two broad priorities for the next decade:
\begin{enumerate}
\item Development of the capability to acquire Doppler measurements of the solar poles, taken from a polar vantage point, and investment in new helioseismic and observational techniques as needed to fully exploit that data.
\item Continued and expanded investment in the Theory/Modeling/Simulation program and in NASA HEC resources.

\end{enumerate}

\polar

\subsection{The Promise of the Poles}

We suggest that \textit{the} most crucial set of observations used to advance understanding of the dynamo in the next decade will be those taken from \textit{a polar vantage point}.    While the polar regions are visible from the ecliptic when the B-angle is favorable, measurements of that region nevertheless suffer from substantial foreshortening.  This effect typically limits the application of local helioseismic analysis to regions that are within 60$^\circ$ of disk center, far from the poles.  When flow measurements are attempted at such extreme latitudes, results obtained using different techniques are in qualitative agreement at best, and only large-scale flows ($\sim$ 100 Mm) are resolved \citep[e.g.,][]{Bogart2015,Hathaway2021}.  This situation is extremely unfortunate because, in many respects, the high latitudes are key to understanding the properties of deep convection.  The reasons for this are fourfold:
\begin{enumerate}
\item Systems with low Ro tend to possess multiple countercells at high-latitudes (Fig. \ref{fig:polar_flow}$a$), whereas those with moderate values of Ro possess no such reversal.   Consensus regarding the high-latitude meridional flow structure will thus provide important clues concerning the nature of deep interior convection, while also placing further constraints on flux-transport models of the dynamo.
\item  The nature of high-latitude differential rotation is similarly sensitive to the amplitude of its underlying convection.  Systems that possess Ro of roughly unity tend to possess polar regions that rotate in prograde fashion with respect to the mid latitudes.  In contrast, systems with low-Ro tend to possess increased spin-down at the poles (Fig. \ref{fig:polar_flow}$b$).  Identifying where the Sun lies on the spectrum between these two extremes will place further constraints on Ro.
\item The vortical motions associated with rotating convection are expected to exhibit their strongest helioseismic signal when viewed from the poles, where the motion is predominantly horizontal at the solar surface.  Measurement of the characteristic spatial size of those structures would place firm constraints on the Rossby Number characteristic of deep convection (e.g., Figs. \ref{fig:rovariation}, \ref{fig:polar_flow}$c,d$).
\item Finally, convection subject to strong rotational influence tends to possess strong latitudinal temperature gradients.  As noted by \citet{Matilsky2020}, the observed structure of solar differential rotation suggests that an equator-to-pole variation of roughly 10 K is expected.  Ecliptic-based measurements, however, indicate at most a weak, if any, variation of  1--2 K \citep{Kuhn1998,Rast2008}.  Emissivity measurements from a direct polar vantage would provide valuable new insight into this puzzle and the nature of deep interior convection.
\end{enumerate}

\noindent\textbf{2.1.1 Nature of Required Polar Observations:}  The observations required to address points 1--4 are long-term Doppler and intensity measurements.  In this context, ``long-term'' does not imply observing throughout a substantial portion of the solar cycle.  The reason is that interior convection, the engine of the dynamo, is expected to operate in roughly steady-state fashion throughout the solar cycle.  In simulations reminiscent of the Sun, convective amplitudes typically do not exhibit strong variation across a magnetic cycle.  The solar differential rotation itself provides a strong observational clue along the same lines;  its relative variation in amplitude across the solar cycle (the so-called torsional oscillations) is on the order of 1\% \citep[e.g.,][]{Howe2009}.

Instead, the need for long-term observations derives from the dominant source of noise affecting both convective- and mean-flow measurements alike:  photospheric supergranulation.  Supergranulation occurs on spatial scales of roughly 30 Mm and amplitudes of a few hundred m s$^{-1}$.  With a lifetime of roughly 1-day, it acts as an intrinsic, stochastic source of noise in any helioseismic velocity determination (regardless of technique or instrument).  The measurement of longer-lived, but much weaker flows requires substantial temporal averaging as a result.   

For the meridional circulation and differential rotation, the time-averaged near-surface profile is required to help address points 1 and 2 above.  For the convection, it is the velocity power spectrum of the convective flows that is of interest, \textit{not} direct imaging of the flows themselves.  Knowledge of where in wavelength space the convective power peaks is sufficient to place strong bounds on Ro.   Both of these features, the near-surface mean flows and subsurface convective spectrum, can be mapped using data from multiple discrete orbital passes.  They do not require continuous observation of the Sun's poles for years at a time.

Ultimately, observations taken across a period of 2 months are sufficient to resolve the spatial spectrum of convective flows with an amplitude of 10 m s$^{-1}$ on spatial scales smaller than supergranulation .   The power spectrum of convective flows with amplitudes of 1 m s$^{-1}$ would require roughly two years of observations.  Similar requirements apply to the meridional circulation and differential rotation \citep{Loptien2015, Hassler2022}.

\subsection{Investment in Theory and Modeling}

\noindent\textbf{A.  Continued and adaptive investment in HPC resources for heliophysics:}    High-performance supercomputing resources are vital to dynamo research, and the resources afforded through NASA's HEC program stand out among peer-level research programs.  It is clear that Heliophysics strives to provide computing time based on project needs.  Many allocations are on-par with what can otherwise only be obtained through the highly competitive and wide-ranging INCITE program run through the Department of Energy.  We make three recommendations regarding NASA Heliophysics HPC support:
\begin{itemize}
 \setlength\itemsep{1pt}
\item Heliophysics and NASA HEC should continue with the current model wherein only a minimal computing time proposal is required for projects that have otherwise been funded by NASA.  This allows a research team to focus on research. 
\item Heliophysics should continue to invest aggressively in HPC infrastructure and resources through NASA HEC and/or separate computing facilities.   This investment should continue to be revisited on an annual basis with input from the community.
\item The ability to modify existing computing allocations throughout the span of their supporting grants should be retained.   Research plans and approaches are often fluid throughout the span of a grant.  Comparable fluidity in the available resources helps ensure such changes do not impact closure on the proposed scientific goals.  
\end{itemize}

\noindent\textbf{B.  Sustained and Expanded Investment in Theory \& Modeling Projects:  }\\We recommend that NASA Heliophysics increase its investment in solar interior projects that primarily entail theory and/or simulation.   This applies to theory and simulation targeted at convection and wave propagation alike  (both were used in the study of \citet{stejko2022}; Fig. \ref{fig:mcvariation}$e$). Theory and simulation are increasingly crucial in the interpretation of existing observations and in the planning of future observations.   A range of specialties are encompassed in the solar-interior focus of heliophysics, but this is not typically true of an individual researcher or even an individual research group. 

Researchers who specialize in theory and simulation lack a regularly-occurring opportunity within NASA Heliophysics to which they can apply.   The Theory/Modeling/Simulation program is competed on a 3-year cycle, but is sometimes supplanted by the mixed-specialization Grand Challenges program.   Other opportunities to study interior flows from a theoretical perspective, within the broader context of observational-based work, such as DRIVE Centers and LWS/FST programs, are similarly competed in sporadic fashion.  

This situation lies in contrast to the Guest Investigator and Supporting Research programs which are competed on an annual basis.   While the latter has supported primarily theoretical investigations in the past, recently, requirements to analyze NASA data have become a normal part of its requirements.  Significant hurdles to sustained, long-term theory-oriented projects arise as a result of these restrictions.  To this end, we make two recommendations:
\begin{itemize}
 \setlength\itemsep{0.01em}
\item Ensure that the current TMS program is retained and competed on a predictable, three-year basis.
\item Create a smaller, theory-focused program that provides funding at a level on par with the existing HSR and HGI programs, and which is competed on an annual basis.
\end{itemize}

%~~~~~~~~~~~~ Bibliography
\newpage
\bibliographystyle{apj_doi_titles}
\bibliography{wp_bib}

\end{document}